# Anomalous diffusion for neuronal growth on surfaces with controlled geometries

Ilya Yurchenko[1], Joao Marcos Vensi Basso[1], Vladyslav Serhiiovych Syrotenko[1], Cristian Staii[1,*]

1. Department of Physics and Astronomy, Center for Nanoscopic Physics, Tufts University, Medford, Massachusetts 02155, *USA*

 * Corresponding author

E-mail: Cristian.Staii@tufts.edu

## ABSTRACT

Geometrical cues are known to play a very important role in neuronal growth and the formation of neuronal networks. Here, we present a detailed analysis of axonal growth and dynamics for neuronal cells cultured on patterned polydimethylsiloxane surfaces. We use fluorescence microscopy to image neurons, quantify their dynamics, and demonstrate that the substrate geometrical patterns cause strong directional alignment of axons. We quantify axonal growth and report a general stochastic approach that quantitatively describes the motion of growth cones. The growth cone dynamics is described by Langevin and Fokker-Planck equations with both deterministic and stochastic contributions. We show that the deterministic terms contain both the angular and speed dependence of axonal growth, and that these two contributions can be separated. Growth alignment is determined by surface geometry, and it is quantified by the deterministic part of the Langevin equation. We combine experimental data with theoretical analysis to measure the key parameters of the growth cone motion: speed and angular distributions, correlation functions, diffusion coefficients, characteristics speeds and damping coefficients. We demonstrate that axonal dynamics displays a cross-over from Brownian motion (Ornstein-Uhlenbeck process) at earlier times to anomalous dynamics (superdiffusion) at later times. The superdiffusive regime is characterized by non-Gaussian speed distributions and power law dependence of the axonal mean square length and the velocity correlation functions. These results demonstrate the importance of geometrical cues in guiding axonal growth, and could lead to new methods for bioengineering novel substrates for controlling neuronal growth and regeneration.

# INTRODUCTION

Neuronal cells are the primary working units of the nervous system. A single neuron is a very specialized cell that develops two types of processes during growth: a long axon and several shorter dendrites (Fig. 1 *a*). These processes extend (grow) and make connections with other neurons thus wiring up the nervous system. Once the neuronal network is formed, a neuron can send electrical signals to other neurons through functional connections (synapses) made between axons and dendrites. During the development of the nervous system axons actively navigate over large distances (~ 10-100 cell diameters in length) to find target dendrites from other neurons and to form neural circuits [1, 2]. Axonal motion is controlled by the growth cone, a dynamic unit located at the leading edge of the axon. The growth cone is sensitive to a great number of external stimuli including biochemical, electrical, mechanical and geometrical cues [1-4].

Many intercellular signaling processes that control growth cone motility have been studied in great detail [2, 5-8], and there is now a substantial amount of information about the molecular pathways that control these processes. For example, there are several models that describe in detail the receptor-ligand interactions, and the changes in the neuron cytoskeleton dynamics in response to biochemical cues from the environment or from other cells [1-6, 8]. However, much less is known about how growth cone dynamics is influenced by external *mechanical* or *geometrical* cues.

Much of the current knowledge about the motility of growth cones comes from *in vitro* studies of neuronal growth on micro-patterned substrates, which is also an area of great importance for studying neuronal regeneration and for bioengineering artificial neural tissue [3, 5, 9-14]. For example, previous experiments have demonstrated that neurons cultured on surfaces with periodic geometrical patterns show a significant increase in the total length of axons, as well as high degree of axonal alignment along certain preferred spatial directions [10, 12, 13, 15-17]. However, many of the previous studies provide mainly qualitative or semi-quantitative descriptions of axonal growth and alignment. A detailed quantitative picture of neuronal growth on surfaces with controlled geometries is still missing.

Quite generally the motion of the growth cone on micro-patterned surfaces is controlled by two main components: a deterministic term leading to axonal growth along certain directions determined by the surface geometry, and a stochastic component leading to random deviations from these growth directions. Stochastic phenomena characterize many processes involved in neuronal growth, including neuron-neuron signaling, fluctuating weak environmental biochemical cues, biochemical reactions taking place in the growth cone, polymerization rates of microtubules and actin filaments, and the formation of lamellipodia and filopodia [1, 2, 6, 18]. Therefore, although the motion of each individual growth cone cannot be predicted, the collective dynamics of ensembles of growth cones belonging to many different neurons can be quantified using stochastic differential equations [12, 19, 20].

In particular, many studies have demonstrated that Langevin and Fokker-Planck equations represent a powerful framework for describing the interplay between the deterministic and stochastic components of biased random motion [9, 12, 19-25]. Theoretical models based on these equations can be used to obtain key dynamical parameters that characterize the cellular motion such as: diffusion (cell motility) coefficients, mean square displacements, velocity and angular correlation functions [19-25]. Moreover, these models provide a systematic approach for analyzing the respective roles played by external biochemical, mechanical, and geometrical cues. For example, in our previous work we have used the Fokker-Planck (F-P) equation to quantify axonal growth on glass [20] and on surfaces with engineered, ratchet-like topography (asymmetric tilted nanorod, or nano-ppx surfaces) [12]. We have demonstrated that axons align along preferred spatial directions on nano-ppx surfaces, and have measured the diffusion coefficient on these substrates. We have shown that axonal alignment originates from the axon-surface interaction forces that produce a "deterministic torque" that tends to align the growth cones along certain preferred growth directions on the surface [12]. In this previous work we have focused on studying the axonal angular distributions, and on measuring the diffusion and angular drift coefficients on the nano-ppx surfaces. The paper contained only a qualitative discussion of the growth dynamics, including time evolution of speed distributions, the velocity autocorrelation functions, and axonal mean square length.

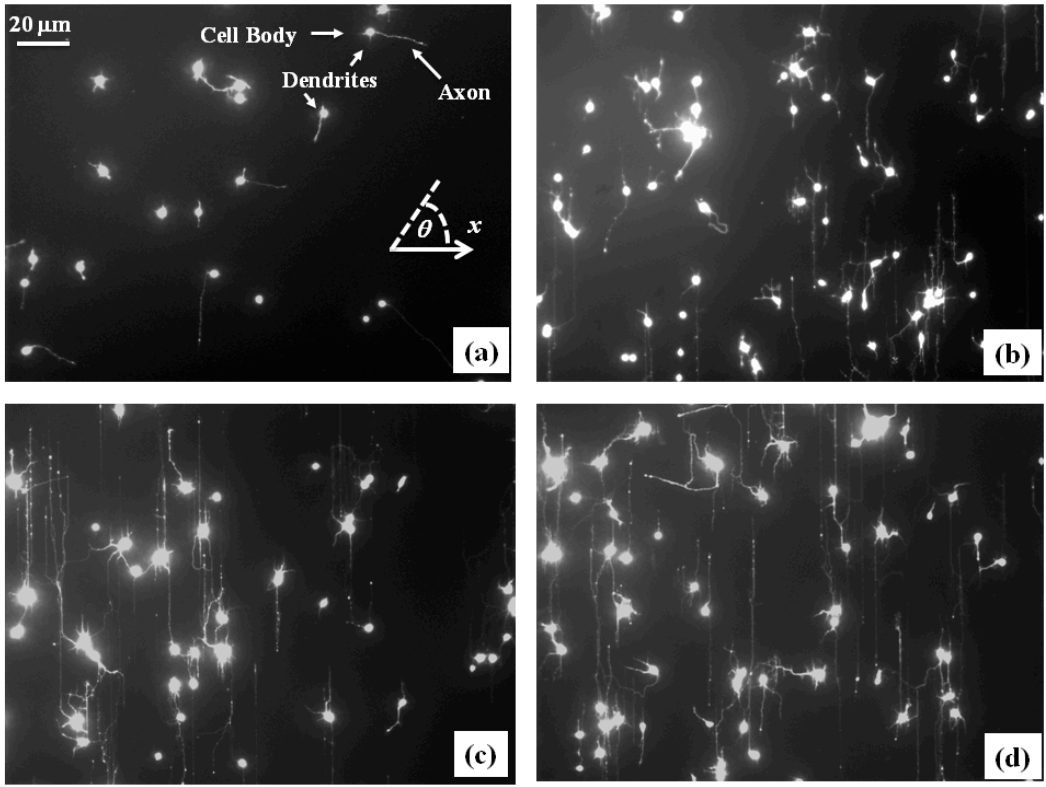

**FIGURE 1. Examples of cultured cortical neurons on PDL coated PDMS surfaces with periodic micro-patterns.** (*a*) Neurons imaged at *t* = 6 hrs after plating. (*b*) Neurons imaged at *t* = 24 hrs after plating. (*c*) Neurons imaged at *t* = 48 hrs after plating. (*d*) Neurons imaged at *t* = 72 hrs after plating. The main structural components of a neuronal cell are labeled in *(a)*. The scale bar shown in (*a*) is the same for all images. The angular coordinate $\theta$ used in this paper is defined in the inset of Fig. 1 *a*. All angles are measured with respect to the *x* axis, defined as the axis perpendicular to the direction of the PDMS patterns (see Fig. 2).

In this paper we present a systematic experimental and theoretical investigation of axonal growth for cortical neurons cultured on poly-D-lysine (PDL) coated polydimethylsiloxane (PDMS) surfaces with periodic micro-patterns (Fig. 1). The periodic geometrical patterns are represented by parallel ridges separated by troughs, with a constant distance between two neighboring ridges $d = 3$ μm (Fig. 2). We demonstrate that axons tend to grow parallel to the surface micro-patterns, and that the degree of axonal alignment is gradually increasing with time. We show that experimental data for axonal growth at early and intermediate times is well-described by linear Langevin equation with Gaussian white noise, i.e. by an Ornstein-Uhlenbeck (OU) process. On the other hand, neuronal growth at longer time scales cannot be described by an OU process. We demonstrate that the growth dynamics at long times is described by superdiffusive dynamics, characterized by non-Gaussian speed distributions and power-law dependence of axonal mean square length. The axonal angular distributions are characterized at all times by Langevin and F-P equations with angular orientation term and stochastic white noise. These models fully account for the experimental data, including growth speeds, axonal alignment, velocity correlation functions, and angular distributions. We extract from the experimental data the main parameters that characterize the motion of the growth cone on PDMS patterned surfaces: diffusion coefficients, velocity correlation functions, damping coefficients, and the deterministic angular orientation terms. These results are important for the fundamental understanding of how geometrical cues influence axonal dynamics, and for bioengineering novel substrates to control neuron growth and regeneration.

**MATERIALS AND METHODS**

**Surface preparation**

The micro-patterns on PDMS surfaces consist of periodic features (parallel ridges separated by troughs). The micro-patterns are characterized by a constant value of the pattern spatial period $d = 3$ μm, defined as the distance between two neighboring ridges (Fig. 2). To obtain these periodic patterns we used a simple fabrication method based on imprinting diffraction grids onto PDMS substrates. We start with 20mL polydimethylsiloxane (PDMS) solution (Silgard, Dow Corning) and pour it over diffraction gratings with slit separations of 3 μm and total surface area 25 x 25 mm$^2$ (Scientrific Pty. and Newport Corp. Irvine, CA). The PDMS films were left to polymerize for 48 hrs at room temperature, then peeled away from the diffraction gratings and cured at $55^0$ C for 3 hrs. We use AFM imaging to ensure that the pattern was successfully transferred from the diffraction grating to the PDMS surface (Fig. 2). The result is a series of periodic patterns (parallel lines with crests and troughs) with constant distance $d = 3$ μm between two adjacent lines (Fig. 2). The surfaces were then glued to glass slides using silicone glue, and dried for 48 hours. Next, each surface was cleaned with sterile water and spin-coated with 3 mL of Poly-D-lysine (PDL) (Sigma-Aldrich, St. Louis, MO) solution of concentration 0.1 mg/mL. The spinning was performed for 10 minutes at 1000 RPM. Prior to cell culture the surfaces have been sterilized using ultraviolet light for 30 minutes.

**Cell culture and plating**

The cells used in this work are cortical neurons obtained from embryonic day 18 rats. For cell dissociation and culture we have used established protocols detailed in our previous work [9, 12, 20, 26, 27]. The brain tissue protocol was approved by Tufts University Institutional Animal Care Use Committee and complies with the NIH guide for the Care and Use of Laboratory Animals. The cortices have been incubated in 5 mL of trypsin at 37ºC for 20 minutes. To inhibit the trypsin we have used 10 mL of soybean trypsin inhibitor (Life Technologies). Next, the neuronal cells have been mechanically dissociated, centrifuged, and the supernatant was removed. After this step the neurons have been re-suspended in 20 mL of neurobasal medium (Life Technologies) enhanced with GlutaMAX, b27 (Life Technologies), and pen/strep. Finally, the neurons have been re-dispersed with a pipette, counted, and plated on PDL coated PDMS substrates, at a density of 5,000 cells/cm$^2$.

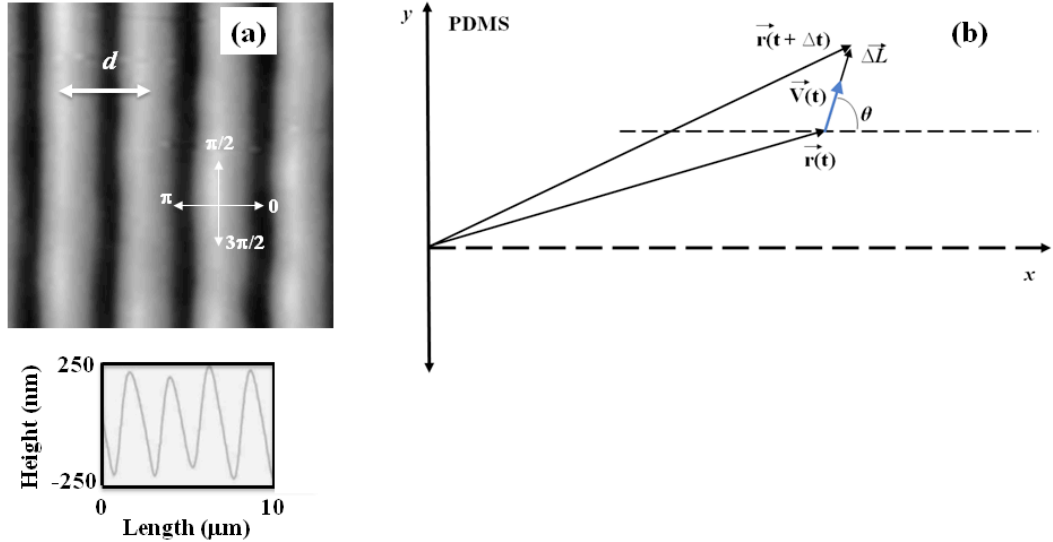

**FIGURE 2. Image of a PDMS surface and definition of the coordinate system.** (a) Topographic Atomic Force Microscope (AFM) image of a PDL coated PDMS patterned surface (top), and example of an AFM line scans obtained across the surface (bottom). *(b)* Coordinate system and the definition of the angular coordinate $\theta$ used in this paper. The *x* axis is defined as the axis perpendicular to the direction of the PDMS patterns. The directions corresponding to $\theta = 0, \pi/2, \pi,$ and $3\pi/2$, and the pattern spatial period *d* (defined as the distance between two neighboring ridges) are also shown in (*a*). The line scan in (*a*) demonstrates that the patterns are periodic in the *x* direction, and have a constant depth of approximately 0.5 μm. The pattern spatial period is *d =3* μm.

**Fluorescence and atomic force microscopy imaging**

Neuronal cells were imaged using an MFP3D atomic force microscope (AFM) equipped with a BioHeater closed fluid cell, and an inverted Nikon Eclipse Ti optical microscope (Micro Video Instruments, Avon, MA). Fluorescence images were acquired using a standard Fluorescein isothiocyanate -FITC filter with excitation of 495 nm, and emission of 521 nm. To acquire the fluorescence images the neurons were incubated for 30 minutes at 37º C with 50 nM Tubulin Tracker Green (Oregon Green 488 Taxol, bis-Acetate, Life Technologies, Grand Island, NY) in phosphate buffered saline (PBS). The samples were then rinsed with PBS and re-immersed in PBS solution for imaging [12, 26]. Fluorescence images were acquired using a standard Fluorescein isothiocyanate -FITC filter: excitation of 495 nm and emission 521 nm. Axon outgrowth was tracked using ImageJ (National Institute of Health). To obtain the angular distributions (Fig. 5 and Fig. S3) all axons have been tracked and then partitioned into segments of 20 μm in length. We have then recorded the angle that each segment makes with the *x* axis (Fig. 2), and the results were plotted as angular histograms (Fig. 5 and S3). The AFM topographical images of the surfaces (Fig. 2) were obtained using the AC mode of operation, and AC 160TS cantilevers (Asylum Research, Santa Barbara, CA). Surfaces were imaged both before and after neuronal culture, and no significant change in topography was observed.

**Data Analysis**

Growth cone position, axonal length, and angular distributions have been measured and quantified using ImageJ (National Institute of Health). The displacement of the growth cone was obtained by measuring the change in the center of the growth cone position. This was determined through image segmentation and tracked using Image J. Axonal growth speed, velocity and angular distributions are quantified at different time points after plating: $t$ = 4, 6, 8, 16, 24, 32, 40, 48, 56, 64, 72, 80, and 88 hrs. The samples were kept in the cell incubator at 37ºC between measurements performed at these different time points. To measure the growth cone velocities the samples were imaged every *Δt = 5 min* for a total period of 30 min for images taken at: $t$ = 4, 6, 8, 16, 24, 32 hrs after plating; 45 min for images taken at 40, 48, 56, 64, and 72 hrs; and for 2 hrs for images taken at 80 and 88 hrs after plating. The increase in total imaging time from 30 min at earlier times to 2 hrs at later times was introduced to accumulate enough statistics for the slower moving growth cones at later time periods, as discussed in the text. The *Δt = 5 min* time interval between measurements was chosen such that the typical displacement $\Delta \vec{L}$ of the growth cone in this interval satisfies two requirements: a) is larger than the experimental precision of our measurement (~ 0.1 μm) [20]; b) the ratio $\Delta \vec{L}/\Delta t$ accurately approximates the instantaneous velocity $\vec{V}$ of the growth cone.

The instantaneous velocity $\vec{V}(t)$ for each growth cone at time *t* is determined by using the formula:

$$\vec{V}(t) = \frac{\Delta \vec{L}}{\Delta t} = \frac{\vec{r}(t + \Delta t) - \vec{r}(t)}{\Delta t} \qquad (1)$$

where $\vec{r}(t)$ is the position vector of the growth cone at time *t*, and $\Delta \vec{L}$ is the net displacement of the same growth cone during the time interval *Δt = 5 min* between the measurements (Fig. 2 *b*). The speed is defined as the magnitude of the velocity vector: $V(t) = |\vec{V}(t)|$, and the growth angle *θ(t)* is measured with respect to the *x* axis (growth angle and the *x* axis are defined in Fig. 2 *b*).

To obtain the speed distributions (Fig. 3 and S2) the range of growth cone speeds at each time point was divided into 15 intervals of equal size $|\Delta \vec{V}_0|$. Experimental data (Fig. 1) shows that over a distance of ~ 20 μm the axons can be approximated by straight line segments, with a high degree of accuracy. Therefore, to obtain the angular distributions (Fig. 5 and Fig. S3) we have tracked all axons using ImageJ and then partitioned them into segments of 20 μm in length. Next, we have recorded the angle that each segment makes with the *x* axis (Fig. 2). The total range [*0, 2π*] of growth angles was divided into 18 intervals of equal size $\Delta \theta_0 = \pi/9$ (Fig 5 and S3).

Experimentally, the velocity autocorrelation function is obtained according to the formula [21, 23]:

$$C_V(t) = \frac{1}{N} \cdot \sum_{i=1}^{N} \left( \vec{V}_i(t) \cdot \vec{V}_i(0) \right) \quad (2)$$

where $N$ is the total number of growth cones and $\vec{V}_i(t)$ represents the velocity of the $i^{th}$ growth cone at time $t$.

**RESULTS**

**Speed distributions, mean square displacement and velocity autocorrelation functions**

The substrates utilized in this study are PDL coated PDMS surfaces, with periodic micro-patterns (parallel ridges separated by troughs, with the pattern spatial period $d = 3$ μm, Fig. 2). The direction of the patterns is shown in Fig. 2 by the parallel white stripes (ridges), as well as by the parallel black stripes (troughs). Cortical neurons are plated on these substrates and axonal growth is quantified at different time points after plating: $t = 4, 6, 8, 16, 24, 32, 40, 48, 56, 64, 72, 80,$ and $88$ hrs. Fig. 1 shows examples of images for axonal growth on these substrates taken at: $t = 6$ hrs (Fig. 1 *a*), $t = 24$ hrs (Fig. 1 *b*), $t = 48$ hrs (Fig. 1 *c*), and $t = 72$ hrs (Fig. 1 *d*). Examples of additional images captured at different times are shown in Fig. S1. The experimental data shows that: a) axons tend to align along the surface patterns, and b) the degree of alignment increases with time. High resolution images (Fig. S4) show that axons tend to grow on top of the ridges.

In this section we analyze the speed distributions and the velocity correlation functions of the growth cones, and measure the dynamical parameters that describe neuronal growth. Fig. 3 shows examples of normalized speed distributions for growth cones that correspond to the images shown in Fig. 1 (examples of additional speed distributions obtained at different times are shown in Fig. S2). To gain a clearer insight into the dynamics that lead to the observed distributions we start with a simple model based on Brownian motion (linear Langevin equation with Gaussian white noise). Models based on Brownian motion (Ornstein – Uhlenbeck process) are extensively used in literature to describe the motility of many different types of cells, including neurons [9, 12, 19-25]. The Ornstein-Uhlenbeck (OU) process is described by the following linear Langevin equation for the speed $V$ [20, 21, 23, 28]:

$$\frac{dV}{dt} = -\gamma_s \cdot (V - V_s) + \Gamma(t) \quad (3)$$

The first term in Eq. 3 represents the deterministic term, and $\gamma_s$ is a constant damping coefficient. The second term $\Gamma(t)$ represents the stochastic change in speed. In the absence of the stochastic term the speed would decay exponentially with a characteristic time: $\tau_s = 1/\gamma_s$, approaching a finite value $V_s$.

We model the stochastic term in Eq. 3 as an uncorrelated Wiener process, satisfying the conditions for Gaussian white noise with zero mean [21, 23, 28]:

$$\langle \Gamma(t) \rangle = 0 \text{ and } \langle \Gamma(t_1) \cdot \Gamma(t_2) \rangle = \sigma^2 \cdot \delta(t_1 - t_2) \quad (4)$$

where $<,>$ denotes the average value, $\sigma^2$ quantifies the strength of the noise, and $\delta(t_1 - t_2)$ is the Dirac delta – function. The Gaussian white noise is a general characteristics of cellular motion, and it reflects the stochastic nature of both the extra cellular (neuron-neuron) signaling [1, 2, 7, 8], as well as the intra-cellular processes, such as: the stochasticity of biochemical reactions taking place in the growth cone, polymerization rates of microtubules and actin filaments, and the formation of lamellipodia and filopodia [1, 4, 19] .

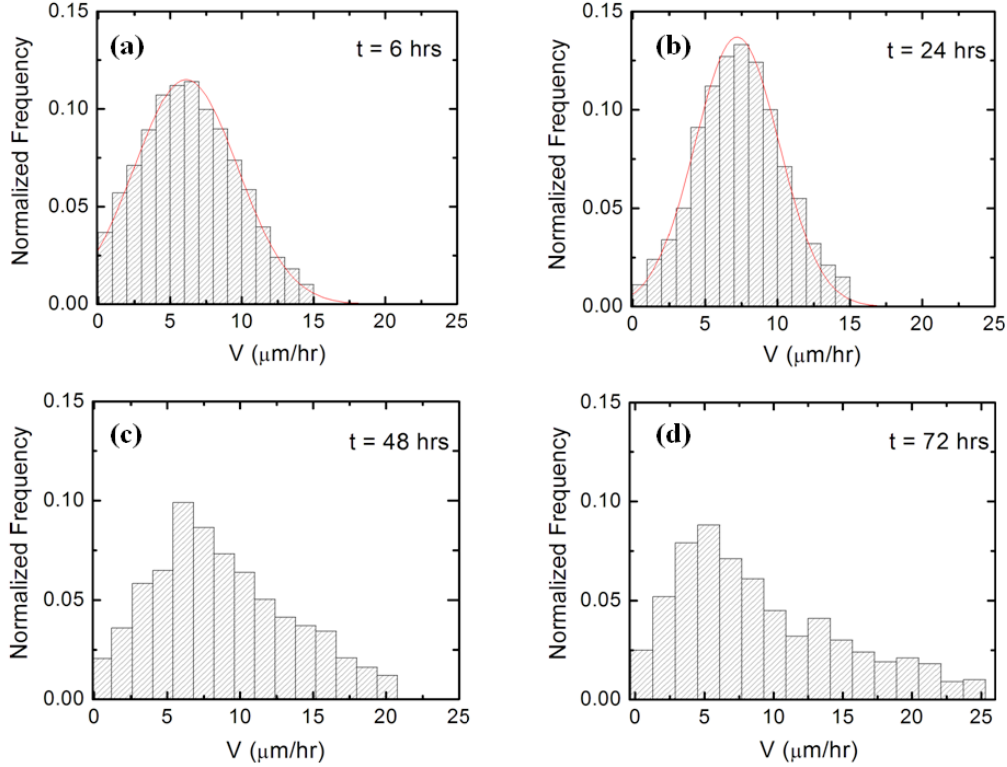

**FIGURE 3. Examples of normalized speed distributions for growth cones measured on PDMS substrates.** (*a*) Speed distribution for N = 168 different growth cones, measured at *t* = 6 hrs after plating. The continuous red curve represents fit with the Gaussian distribution given by Eq. 6. (*b*) Speed distribution for N = 189 different growth cones measured at *t* = 24 hrs after plating. The continuous red curve represent fit with the Gaussian distribution given by Eq. 6. (*c*) Speed distribution for N = 176 growth different cones measured at *t* = 48 hrs after plating. (*d*) Speed distribution for N = 192 different growth cones measured at *t* = 72 hrs after plating.

To obtain the speed distributions $p(V,t)$ of axonal growth we write the Fokker-Planck equation corresponding to Langevin Eq. 3 [21, 23, 28]:

$$\frac{\partial p(V,t)}{\partial t} = \frac{\partial}{\partial V}[\gamma_S \cdot (V - V_S) \cdot p(V,t)] + \frac{1}{2}\sigma^2 \cdot \frac{\partial^2 p(V,t)}{\partial V^2} \quad (5)$$

The stationary solution of Eq. 5 is given by [21, 28]:

$$p(V) = p_0 \cdot \exp\left(-\frac{\gamma_S}{\sigma^2} \cdot (V - V_S)^2\right) \quad (6)$$

where $p_0$ is a normalization constant obtained from the condition: $\int_0^\infty p(V)\,dV = 1$. The probability distribution given by Eq. 6 can be approximated by a Gaussian as long as the negative tail of the Gaussian is smaller than the width of the distribution: $\int_{-\infty}^{0} p(V) dV \ll \sqrt{\frac{\sigma^2 \cdot \pi}{\gamma_s}}$.

This condition is indeed satisfied by our data. Fig. 3 *a*, *b* and Fig. S2 *a*, *b* show examples of experimental data for the growth cone speed distribution, together with the fit of the data (continuous red curve) obtained by using Eq. 6. We conclude that the speed distribution data for $t < 48$ hrs is well described by normalized Gaussian functions. However, this is not the case for growth at later times: $t \geq 48$ hrs. For example, Fig. 3 *c*, *d* and Fig. S2 *c*, *d* show distributions with long tails at high speeds, which display a significantly greater number of high speed events than expected from the Gaussian distribution of Eq. 6. Thus the experimental data indicates that the long-term motion of the axons cannot be described by the simple OU (diffusive) behavior.

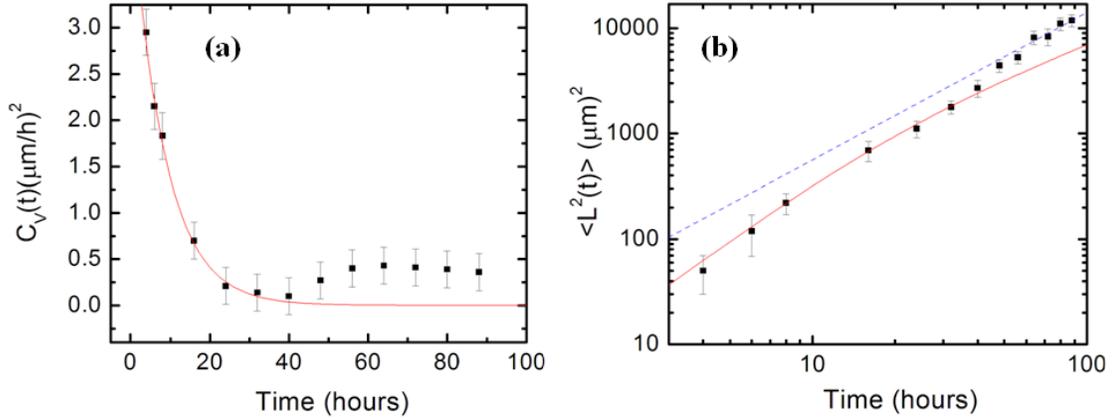

**FIGURE 4. Variation of the velocity autocorrelation function and axonal mean square length with time.** (*a*) Data points: experimentally measured velocity autocorrelation function vs. time. The continuous red curve represents the fit of the data points measured for $t < 48$ hrs with the prediction of the theoretical model based on the Ornstein-Uhlenbeck process (Eq. 7). *(b)* log-log plot of axonal mean square length vs. time. The continuous red curve represents the fit to the data measured at $t < 48$ hrs with Eq. 8 (prediction of the theoretical model based on the Ornstein-Uhlenbeck process). The dotted blue curve represents the fit to the data points measured for $t \geq 48$ hrs with a power-law function (Eq. 11 and Eq 15). Each data point in (*a*) and (*b*) was obtained by measuring between N = 150 and N = 195 different axons (corresponding to 5-10 different fluorescent images per time data point). Error bars in both figures indicate the standard error of the mean. The fit of the data in with Eq. 7 for (*a*), and Eq. 8 for (*b*) give the diffusion coefficient *D* and the constant damping coefficient $\gamma$ of the Ornstein-Uhlenbeck process (see text).

To further investigate the dynamics of the growth cones on micro-patterned PDMS surfaces we measure the velocity correlation functions and the axonal mean square lengths as a function of time (Fig. 4). Since the axonal motion takes place in two spatial dimensions, the OU process implies the following expressions for the growth cone velocity autocorrelation function $C_V(t)$, and for the axonal mean square length $<\vec{L}^2(t)>$ as functions of time [21, 23, 25]:

$$C_V(t) \equiv \langle \vec{V}(t) \cdot \vec{V}(0) \rangle = 2D \cdot \gamma \cdot e^{-\gamma t} \qquad (7)$$

$$\langle \vec{L}^2(t) \rangle = 4D \cdot t - \frac{4D}{\gamma} \cdot (1 - e^{-\gamma t}) \tag{8}$$

where $\gamma$ is the constant damping coefficient that characterizes the exponential decay of the autocorrelation functions, and $D$ is the cell random motility coefficient [1, 20, 23-25]. At the level of cell populations the random motility coefficient is analogous to the diffusion coefficient of the OU process [23, 25]. In this paper we will refer to $D$ as diffusion coefficient, as it is customary in literature [1, 20-25].

Fig. 4 *a* shows the experimental data for the velocity autocorrelation function vs. time, together with the fit of the data (continuous red curve) with Eq. 7, which represents the theoretical prediction based on the OU process. Fig. 4 *b* shows the experimental data for the axonal mean square length vs. time, together with the fit of the data (continuous red curve) obtained by using Eq. 8. Brownian motion described in terms of the OU dynamics is characterized by: 1) exponential decay of the velocity autocorrelation function with time *t*, and 2) mean square displacement (axonal mean square length) proportional to $t^2$ for short times, and linear increase with *t* for longer times (normal diffusion). These predictions are represented by the continuous red curves in Fig. 4. (the dotted blue line in Fig. 4 *b* represents fit with a power-law function as explained below). However, the plots in Fig. 4 *a* and Fig. 4 *b* show two distinct regimes. For small/intermediate time scales $t < 48$ hrs the experimental data is well described by the OU model, while for longer time scales $t \geq 48$ hrs the experimental data differ significantly from the predictions of the OU process.

We focus first on analyzing the data measured at small/intermediate times. From the fit of the data in Fig. 4 with Eqs. 7 and 8 we obtain the following values for the diffusion coefficient $D$ and for the constant damping coefficient $\gamma$ : $D = (19 \pm 2)\,\mu m^2/hr$ and $\gamma = (0.12 \pm 0.06)\,hr^{-1}$ (both values obtained for $t < 48$ hrs). The value for the diffusion coefficient is close to the diffusion coefficients we have obtained in previous work for neuronal growth on glass and nano-ppx surfaces [12, 20] . From the value of the damping coefficient we find a characteristic time for the exponential decay of the velocity autocorrelation function: $\tau = 1/\gamma \approx 8\,hr$.

Since the axonal growth for $t < 48$ hrs is described by an OU process, we can relate $D$ and $\gamma$ with a typical mean square velocity for neuronal growth on PDMS surfaces via the general expression (valid for any OU process) [21, 23, 25]:

$$D = \frac{\langle V_c^2 \rangle \cdot \tau}{2} = \frac{\langle V_c^2 \rangle}{2\gamma} \tag{9}$$

Using the experimentally measured values for $D$ and $\gamma$, Eq. 9 predicts the following value for the characteristic speed of neuronal growth on micro-patterned PDMS surfaces:

$$V_c \equiv \sqrt{\langle V_c^2 \rangle} = \sqrt{2D\gamma} \approx 2.1\,\mu m/hr \tag{10}$$

This value is in good agreement with the root mean square (rms) of the speed distributions showed in Fig. 3 *a* and *b*.

In conclusion, we found that neuronal growth on micro-patterned PDMS substrates is well-described by an Ornstein-Uhlenbeck process (linear Langevin equation with Gaussian white noise) at low and intermediate growth times. By fitting the experimental data with the theoretical OU model (Fig. 4) we measure the fundamental dynamical parameters for neuronal growth on these substrates: diffusion coefficient $D$, constant damping coefficient $\gamma$, characteristic time $\tau$, and use these values to calculate a characteristic speed of axonal growth $V_c$. These values are comparable with the corresponding values we have previously obtained by using the Fokker-Planck equation for describing neuronal growth on glass and nano-ppx surfaces [12, 20].

**Anomalous growth at long time scales**

The experimental data presented in Fig. 4 shows a transition between an initial OU process and anomalous dynamics at longer time scales. The velocity autocorrelation function in Fig. 4 *a* displays a gradual transition between an exponential decay for $t < 48$ hrs (red continuous curve) and a slower time dependence for $t \geq 48$ hrs. Furthermore, the double-logarithmic plot of $<\vec{L}^2(t)>$ vs. time in Fig. 4 *b*, shows a crossover between quadratic (i.e. ballistic) and linear (diffusive) time dependence at small and intermediate *t* and a power law behavior at longer times. This crossover takes place within the same time scale as the velocity autocorrelation function $C_V(t)$ (Fig. 4 *a*). Finally, the speed distributions in Fig. 3 *c, d* (as well as in Fig. S2 *c, d*) clearly display non-Gaussian probability distributions in contrast with the predictions of the OU model. All these experimental findings are signatures of anomalous diffusion as we will discuss below.

The dotted blue line in Fig. 4 *b* represents the resulting fit of the axonal mean square length with a power-law function:

$$<\vec{L}^2(t)> \sim t^\mu \quad (11)$$

From the fit we obtain: $\mu = 1.4 \pm 0.2$. Thus Eq. 11 is describing superdiffusive dynamics (power law with exponent $> 1$) for axonal growth at long time scales: $t \geq 48$ hrs [32, 33]. We note that the superdiffusion is ceasing once all the axons meet dendrites (or other axons from different neurons), and form connections. At that point in time ($t > 90$ hrs in our experiments) the overall motion of the growth cones comes to an end, and all the growth speeds and velocity correlation functions are equal to zero. We will present the significance of the observed superdiffusive behavior in the Discussion section below.

**Axonal alignment with the surface patterns and dynamics of angular growth**

To gain further insight into the growth cone dynamics we analyze the angular distributions of axons on micro-patterned PDMS surfaces. Images of axonal growth on these surfaces and the corresponding angular distributions measured at different time points are shown in Fig. 1, and respectively in Fig. 5 (additional images and angular distributions are presented in Fig. S1 and S3). The experimental data demonstrates that axons tend to align with the surface patterns (peaks at $\theta = \pi/2$ and $\theta = 3\pi/2$ respectively), and that degree of alignment is increasing with time.

In our previous work [12] we have used linear Langevin and F-P equations to describe the angular dynamics of the growth cones on nano-ppx surfaces. A similar formalism has been used by other authors, for example to describe motion of human granulocyte cells under the influence of electric fields [21], or the migration of microvessel endothelial cells [25]. Using the growth angle $\theta(t)$ defined in Fig 1 *a* and Fig 2 one obtains the following Langevin and F-P equations, for the angular motion of axons on micro-patterned PDMS substrates [12, 21, 28]:

$$\frac{d\theta}{dt} = -\gamma_\theta \cdot \cos\theta(t) + \Gamma_\theta(t) \quad (12)$$

$$\frac{\partial p(\theta,t)}{\partial t} = \frac{\partial}{\partial \theta}\left[-\gamma_\theta \cdot \cos\theta(t) \cdot p(\theta,t)\right] + D_\theta \cdot \frac{\partial^2 p(\theta,t)}{\partial \theta^2} \tag{13}$$

where $p(\theta,t)$ is the probability distribution for growth angles, $D_\theta$ is an effective *angular* diffusion (cell motility) coefficient, and $\Gamma_\theta(t)$ is the stochastic change in angle. As in ref. [12] the term $-\gamma_\theta \cdot \cos\theta(t)$ on the right hand side of Eq. 12 and 13 corresponds to a "deterministic torque", which represents the tendency of the growth cone to align with the preferred growth direction imposed by the surface geometry. We note that this term has a maximum value if the growth cone moves perpendicular to the surface patterns ($\theta = 0$ or $\theta = \pi$), and the cell-surface interaction tend to align the axon with the surface pattern. Moreover, as we will discuss below, the strength of the interaction between the patterned surface and the growth cone is quantified by the magnitude of the torque $\gamma_\theta$.

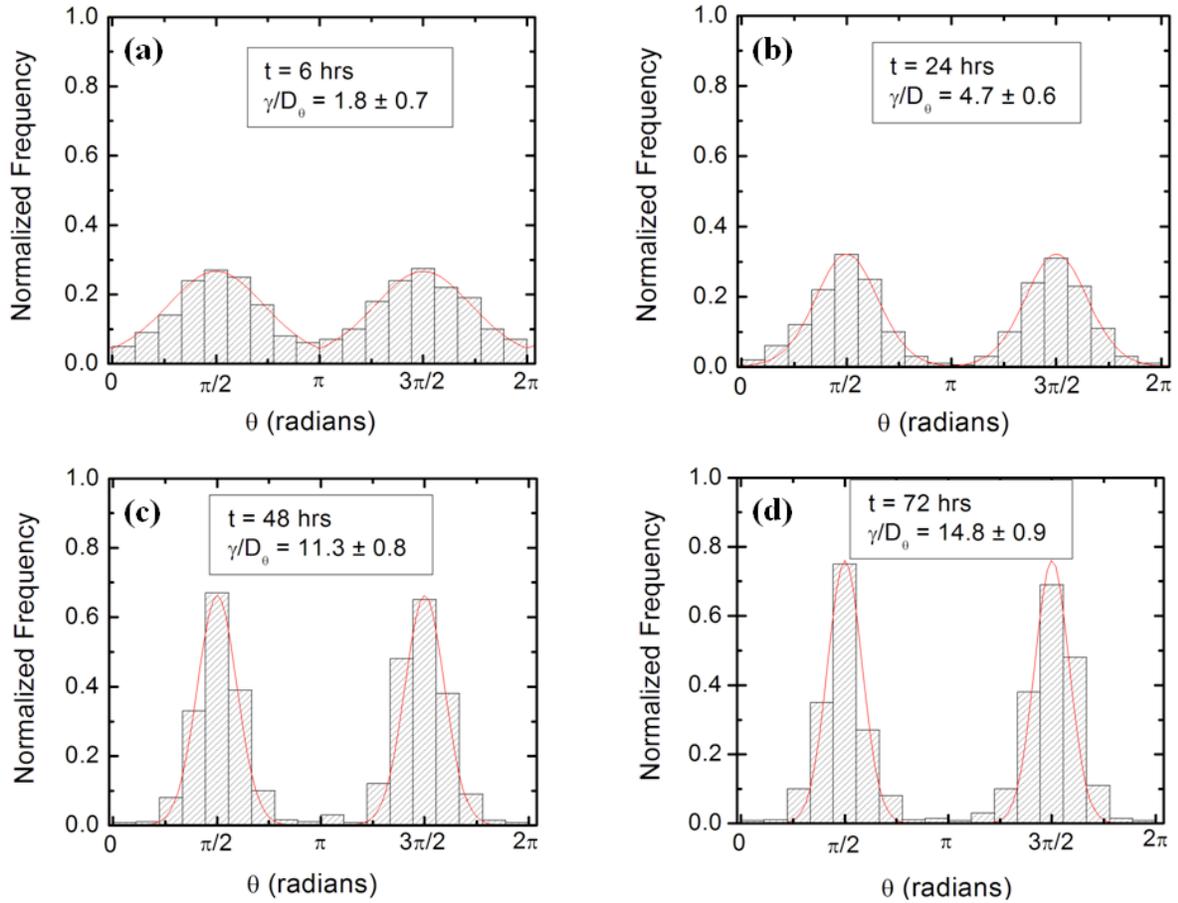

**FIGURE 5. Examples of normalized experimental angular distributions for axonal growth.** The vertical axis (labeled Normalized Frequency) represents the ratio between the number of axonal segments growing in a given direction and the total number N of axon segments measured at a given time $t$. Each axonal segment is of 20 μm in length (see Data Analysis section). (*a*) Data for N = 1724 different axon segments obtained at $t$ = 6 hrs after plating. (*b*) Data for N = 2078 different axon segments obtained at $t$ = 24 hrs after plating. (*c*) Data for N = 2405 different axon segments obtained at $t$ = 48 hrs after plating. (*d*) Data for N = 2629 different axon segments obtained at $t$ = 72 hrs after plating. The data shows that the axons

display strong directional alignment along the surface patterns (peaks at $\theta = \pi/2$ and $\theta = 3\pi/2$), with the degree of alignment (sharpness of the distribution) increasing with time. The continuous red curves in each figure represent fit to the data points using Eq. 14. The fit gives the ratio $\gamma_\theta / D_\theta$ between the deterministic torque and the diffusion coefficient for the angular motion, at each time point (see text).

The stationary solution of Eq. 13 at each time point is given by [12, 21, 28]:

$$p(\theta) = A_0 \cdot \exp\left( \frac{\gamma_\theta}{D_\theta} \cdot |\sin(\theta)| \right) \quad (14)$$

where $A$ is a normalization constant.

The absolute value $|\sin \theta|$ reflects the symmetry of the growth around the $x$ axis: the two distributions centered at $\theta = \pi/2$ and $\theta = 3\pi/2$ are symmetric with respect to the directions $\theta = \pi$ and $\theta = 0$, (as shown in Fig. 5 and Fig. S3), which in turn means that there is no preferred direction along the pattern (i.e. the "up" and "down" directions in Fig. 1 and Fig. 2 are equivalent for neuronal growth). This conclusion holds for each time point considered in these experiments. We note the difference between these results and our previous results obtained on directional nano-ppx surfaces, where we have reported an additional unidirectional bias beyond the preferential alignment along the surface patterns [12].

We use Eq. 14 to fit the normalized experimental angular distributions at each time point considered in these experiments (fits to the data are represented by the continuous red curves in Fig. 5 and Fig. S3). Eq. 14 shows that the angular distributions give only the *ratios* $\gamma_\theta / D_\theta$, between the deterministic torque and the angular diffusion coefficient, measured at each time point. Fig 6 shows the variation of this ratio with time, for all times considered in this experiment.

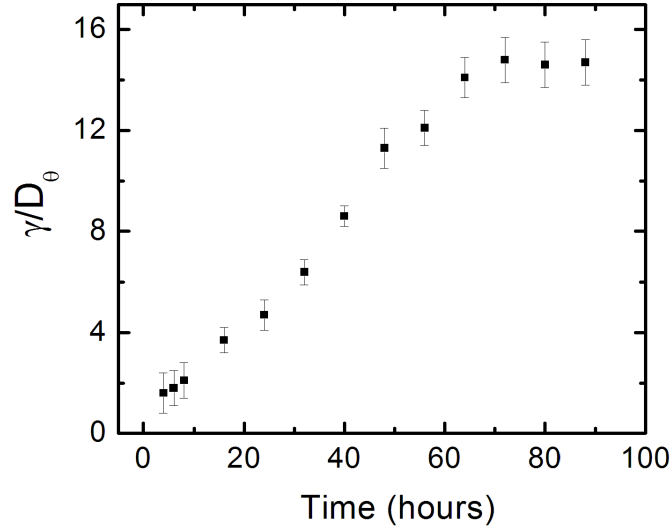

**FIGURE 6. Variation in time of the ratio between the deterministic torque and the diffusion coefficient for the angular motion.** The increase in the ratio $\gamma_\theta / D_\theta$ reflects the increase in the cell-surface interactions as discussed in the text. Error bars indicate the uncertainties obtained from the fit of the normalized angular distributions (Fig. 5).

**Neuronal growth on flat PDMS substrates**

We have also performed controlled experiments of neuronal growth on flat (un-patterned) PDMS substrates. Images of neuronal growth were captured at different time points after plating (examples of neuronal growth on these substrates are shown in Fig. S5). These controlled experiments show no axonal alignment, and flat angular distributions (Fig. S5 and Fig. S6). The corresponding velocity correlation functions and the axonal mean square lengths as a function of time are shown in Fig. S7. The experimental data demonstrates that the dynamics on flat PDMS surfaces is well described by OU dynamics at all time points. In particular, in contrast with the results obtained for patterned PDMS substrates, there is no cross-over between normal and superdiffusive behavior for neurons cultured on flat PDMS surfaces. We have performed an analysis of velocity autocorrelation data and axonal mean square length for neurons cultured on flat PDMS substrates, which is similar to the analysis of the corresponding data for patterned PDMS (Eq. 7 and 8). We obtain the following values for the diffusion coefficient $D$ and for the constant damping coefficient $\gamma$: $D = (17 \pm 3)\, \mu m^2/hr$ and $\gamma = (0.1 \pm 0.07)\, hr^{-1}$. These values characterize the axonal dynamics at all time points on un-patterned PDMS, and are close to the corresponding values obtained for the OU dynamics at short/intermediate time scales on patterned PDMS substrates (see previous sections).

**DISCUSSION**

The experimental data for speed distributions (Fig. 3 and Fig. S2), as well as velocity autocorrelation functions and mean square length (Fig. 4 and Fig. S3) for axons grown on micro-patterned PDMS surfaces show a gradual cross over between normal diffusion (described by the OU process), observed for low/intermediate time scales ($t < 48$ hrs), and superdiffusion observed for larger time scales ($t \geq 48$ hrs). In contrast, experiments performed on flat PDMS substrates show normal diffusion at all times. The OU process, which is inspired by the study of the Brownian motion, represents the simplest stochastic model used for describing cellular motility. It has been successfully used for modeling the dynamics of many types of cells including endothelial cells [25], human granulocytes [21], fibroblasts and human keratinocytes [23], as well as cortical neurons [12, 19, 20]. The values we have obtained for the diffusion coefficient $D = (19 \pm 2)\, \mu m^2/hr$ and the characteristic rms speed for growth cones on PDMS surfaces (Eq. 10) are comparable with the corresponding values reported for human peritoneal mesothelial cells [29], one order of magnitude smaller than the values reported for human keratinocytes [23], and for endothelial cells [25], and about two orders of magnitude smaller than the corresponding values reported for glioma cells [30]. These results are consistent with the relatively slower dynamics expected for growth cones as they move to form connections and to wire up the nervous system [1, 2].

Anomalous diffusion has been reported in literature for a large number of physical systems including: charge transport in semiconductors [34], reptation dynamics in polymers [35], fluid dynamics [36, 37], quantum optics [38], and even bird flight [33, 39]. More recently, superdiffusion has been observed in both eukaryotic [32] and bacterial [33] cellular motion. Anomalous diffusion has been modeled in several ways, including generalized Langevin equations, generalized master equations, and fractional diffusion equations (for a general review article see ref. [36]). In particular equations with fractional derivatives, such as the fractional Klein-Kramers equation, have been used to describe Lévy-flight type diffusion processes. For example, Dietrich and collaborators have used the fractional Klein-Kramers equation to describe the superdiffusive dynamics of transformed canine kidney cells [32]. Their data shows transition between different superdiffusion regimes for the motion of these cells, which happens at much lower time scales (order of 10 - 100 minutes) than in our experiments. One of the main advantages of the fractional Klein-Kramers models is that they provide, in principle, a unified framework for describing the dynamics of the diffusion process. These models include the OU process as a special limiting case (fractional exponent equals 1) [32, 36]. However, we note that the solutions of the fractional Klein-Kramer equation are only known in general for time scales larger than the characteristic decay time of the superdiffusive process [32, 36]. In this limit, the solution for the mean square displacement (mean axonal length) is given by [32, 36]:

$$<\vec{L}^2(t)> \approx \frac{2D_\alpha}{\Gamma_1(3-\alpha)} t^{2-\alpha} \qquad (15)$$

where $\Gamma_1$ is the mathematical gamma function (not to be confounded with the stochastic terms in Langevin Eq. 3 and 12), and $D_\alpha$ is the generalized diffusion coefficient. The limiting case $\alpha = 1$ for the exponent in the above equation describes normal diffusion (OU process). A fit with the power-law model given by Eq. 15 (represented by the dotted blue line in Fig. 4 *b*) gives the following parameters for the anomalous dynamics of neuronal growth on micro-pattern PDMS surfaces: exponent $\alpha = 0.6 \pm 0.2$, and the generalized diffusion coefficient $D_\alpha \approx 14 \mu m^2/hr$. This value for $\alpha$ gives the superdiffusive behavior $<\vec{L}^2(t)> \sim t^{1.4}$ of Eq. 11.

It has been suggested in literature [32, 33, 36, 44] that the transition between the normal and superdiffusive behavior in the case of cellular motion reflects the emergence of long-range temporal and spatial correlations in the underlying dynamics. The cross over time between these two regimes depends on the cell type. For example, in the case of transformed canine kidney cells it was found that this time is of the order ~ 1-2 hr [32], and similar time scales have also been reported for endothermal Hydra cells [44]. This is one order of magnitude lower than the cross-over time found in our experiments t ~ 48 hrs.

The difference in cross-over times between neurons and other types of cells can be qualitatively explained, by considering the characteristic features of axonal growth. First, we note that the growth cones exhibit lower motility compared to other types of cells, such as human keratinocytes, endothelial [25], or glioma cells [30]. There are several factors that contribute to the growth cone motion, including the formation of adhesive contact points, changes in the internal dynamics of the cytoskeleton, coupling between receptors and cytoskeletal components, activation of signaling pathways, and the generation of traction forces through the activity of myosin II [1,2,5]. Some of these guidance mechanisms are common to many other different cell types, others are specific to neurons. For example, the growth cones move forward and steer by using a "clutch mechanism" in which the receptor binding to the substrate leads to the formation of a complex that couples the receptors and the actin flow and controls the extension of the filopodia [1]. For the growth cone to accurately follow geometrical cues, this motility machinery must have the potential to be biased asymmetrically, and to achieve accurate steering and turning that ultimately leads to the formation of complex neuronal networks. In addition, a growth cone is also able to modulate its response as it moves on the substrate. A growing body of work indicates that geometrical and mechanical cues can also affect, either directly or indirectly, the transcriptional regulation of mechanosensing proteins involved in embryogenesis and pattern formation [45].

A detailed understanding of how all these multiple processes work together to achieve the accurate steering and turning of the growth cone is still lacking. However, it is known that the time frame for developing this complex axonal migration machinery is of the order of tens of hours to several days [1, 2, 6, 8], depending on the external conditions. This time frame is consistent with our finding of t ~ 48 hrs for the cross-over time between normal and superdiffusive dynamics. We emphasize that this is time frame is measured for neuronal growth in controlled geometries, where the growth cone motion is not random, and it is directed by the surface geometrical features. We hypothesize that the long range correlations which give rise to the observed superdiffusive dynamics represent a measure of the neuron-surface interactions. This hypothesis can be tested in future studies that use a combination of traction force microscopy to quantify cell-surface interactions, and fluorescence techniques to measure the density of cell-surface receptors (see below).

The above conclusions are also supported by our analysis of the angular motility of the growth cones. In this case, the angular distributions are described by Eq. 14 (the solution of the Fokker-Planck Eq. 13). We use Eq. 14 to fit the experimental data and to quantify the variation with time of the ratio $\gamma_\theta / D_\theta$ between cell-surface coupling torque and the angular diffusion coefficient. Fig. 6 shows this variation for all the time data points considered in our experiments. We note that the magnitude of this ratio is increasing with time, which leads to narrower angular distributions for later times, i.e. higher degree of alignment. The increase in the $\gamma_\theta / D_\theta$ ratio means either an increase in the magnitude of the deterministic torque $\gamma_\theta$, a decrease in the coefficient of angular diffusion $D_\theta$, or both of these processes taking place simultaneously. We note that all these cases imply that the strength of the interaction that tends to align the axons to the patterned lines on PDMS increases with time. A larger value for $\gamma_\theta$ means stronger tendency for alignment for all angles $\theta \neq \pi/2$, and $\theta \neq 3\pi/2$, i.e. for all growth directions which are not parallel with the surface patterns (Eq. 12). The value of $\gamma_\theta$ has no effect on the motion of the growth cone once the axon is aligned with the surface pattern ($\theta = \pi/2, 3\pi/2$), i.e. the overall torque in this case is zero and the growth cone continues to move along the pattern. Consequently, once aligned with the patterns the axons continue to grow preferentially along the direction of the patterns, and the coefficients of angular diffusion becomes smaller. That is, a larger degree of alignment between the axon and the surface pattern, results in smaller values for $D_\theta$. We compare the values for the ratios $\gamma_\theta / D_\theta$ measured for neuronal growth on PDMS surfaces with the corresponding values obtained for growth on nano-ppx surfaces [12]. On nano-ppx we found these ratios in the range: $2 < \gamma_\theta / D_\theta < 8$ [12]. Fig. 6 shows that on micro-patterned PDMS surfaces this ratio varies between ~ 2 ($t$ = 6 hrs) to ~ 15 ($t$ = 88 hrs), which is compatible with a larger degree of axonal alignment measured on PDMS compared to the nano-ppx substrates.

We interpret these experimental results as follows. Initially, when the axons start growing, the growth cone moves through the surrounding environment by executing a random walk (OU process) on the surface [1, 2 19, 20]. At later times, the interactions between the neurons and the patterned PDMS substrates (cell-surface forces and deterministic torque) tend to align the motion of the growth cone along the preferred directions determined by the surface patterns. Therefore the growth cone tends to rotate as it extends and aligns with the patterns, which leads to an increasing degree of axonal alignment with time. We emphasize that the directional motion of the growth cone results from *two* combined effects: a) growth cones are more likely to move in the directions parallel with the surface patterns (cosine dependence in Eq.

12 tends to rotate the growth cone along these directions); and b) growth cones that are moving along the direction of the surface patterns exhibit superdiffusive dynamics at later times (Eq. 17). It is the combination of these two effects that is ultimately responsible for the high degree of axonal alignment and long axonal lengths observed in our experiments.

These results are consistent with contact-guidance behavior that we and other groups have previously reported for neurons grown on different types of substrates [12, 17, 31, 40, 41]. Contact guidance is the phenomenon in which cells orient their motion in response to surface geometrical cues. This behavior has been reported for many types of cells including granulocytes, fibroblasts, and tumor cells [31, 40-42]. Our experiments demonstrate contact-guidance behavior for cortical neurons grown on substrates with controlled geometries. Specifically, we show that axons grow preferentially in the direction of micro-patterned parallel lines on PDMS surfaces. The pattern spatial period ($d = 3$ μm) is comparable with the dimensions of the growth cones. Growth cones have several different types of surface receptors and membrane curvature sensing proteins involved in surface adhesion, and locomotion including amphipathic helices and bin-amphiphysin-rvs (BAR) - domain containing proteins [1, 5, 31]. Previous studies have shown that, in the case of contact-guidance, an increase in the density of anchored surface receptors leads to a higher degree of directional cell motility [17, 31, 40-42]. An important parameter for contact guidance is the ratio between the size of the growth cone and the characteristic dimensions of the surface geometrical features [31]. This parameter determines the surface density of surface receptors, which mediate adhesion and mechanotransduction between the cell cytoskeleton and the substrate. We hypothesize that for neurons grown on PDMS surfaces where the linear dimension of the growth cone matches the pattern spatial period, the growth cone "wraps tightly" around the surface features, which results in a minimum contact area and thus maximum density of surface receptors. Previous reports have shown that the maturation of the surface receptor and focal adhesion points respond to external forces, including cell-substrate traction forces [43]. Thus high-curvature features such as ridges (see AFM images of the PDMS patterns in Fig. 2) will impart higher forces to the focal contacts of filopodia wrapped over these features, compared to those contacting low-curvature patterns. Furthermore, microtubules and actin filaments inside the growth cone act as stiff load-bearing structures that provide resistive forces to the bending of the filopodia. Together these effects will ultimately lead to axonal alignment along the PDMS surface, as reported in this paper.

We have shown that the theoretical model based on Langevin and F-P equations: 1) fully accounts for the experimental data of neuronal growth on PDMS surfaces; 2) has a minimum number of phenomenological parameters that account for the cell-surface interactions; and 3) allows for meaningful comparisons between different growth dynamics that change with time, as well as for comparisons with the simpler case of linear Langevin dynamics that describes neuronal dynamics on glass, and with the dynamics of other types of cells reported in literature. The model predicts characteristic speeds for neuronal growth and deterministic torque that tends to align axons along certain preferred directions along the surface, and it describes cross over between linear (OU) and anomalous dynamics. We hypothesize that these could be general features of cellular motility in various environments with inhomogeneous physical and chemical properties. Evidence in support of this hypothesis comes from previous studies of neuronal growth on surfaces with various geometries, textures and biochemical properties [3, 7-20], as well as from motility studies for other types of cells [21-25]. In addition, the model could be further extended to account for the explicit dependence of the phenomenological parameters on the surface geometrical properties (such as pattern period *d* for the PDMS surfaces presented here). This will require measuring cell-surface coupling forces (using for e.g. traction force microscopy) and quantifying the density of cell surface receptors (using fluorescence techniques) that determine axonal contact guidance dynamics. In principle these future studies will enable to quantify the influence of environmental cues (geometrical, mechanical, biochemical) on neuronal growth, and to correlate the observed growth dynamics with cellular processes (cytoskeleton dynamics, cell-surface interactions, cell-cell communication etc.).

 **CONCULSIONS**

In this paper, we have used stochastic analysis to model neuronal growth on micro-patterned PDMS substrates coated with PDL. We have shown that the experimental data for small and intermediate time scales are well-described by Ornstein-Uhlenbeck (OU) processes (linear Langevin equations with white noise). On the other hand, growth measured at longer time scales displays superdiffusive dynamics, characterized by non-Gaussian speed distributions, and power-law behavior of velocity autocorrelation function and of the axonal mean square length.
These results are consistent with contact-guidance phenomenon for neuronal growth, and imply the existence of long-range correlations of the underlying dynamics, which are imparted by the surface geometry. Our approach offers a general theoretical framework that could be applied to neurons cultured on other types of substrates with different geometrical features as well as to neuronal growth *in vivo*. Moreover this model could be applied to the motion of other types of cells in controlled environments including electric fields, surfaces with different stiffness, or biomolecular cues with different concentration gradients.

**ACKNOWLEDGEMENTS**

The authors thank Prof. David Kaplan's laboratory at Tufts Biomedical Engineering for providing embryonic rat brain tissues. The authors gratefully acknowledge financial support for this work from Tufts Summer Scholars (IY), and Tufts Faculty Award (FRAC) (JMVB, CS).

**Supporting information**

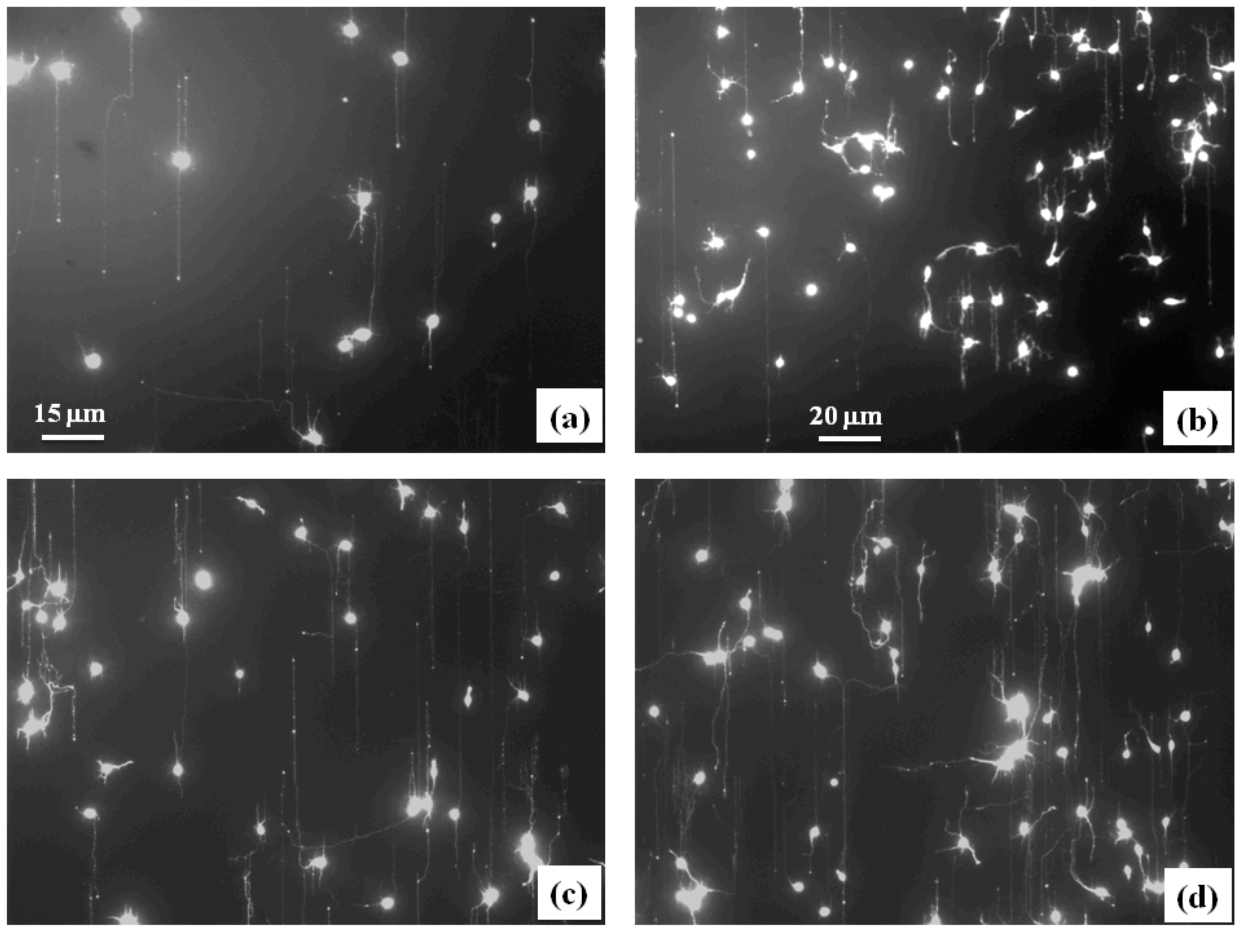

**FIGURE S1. Examples of cultured cortical neurons on PDL coated PDMS surfaces with periodic micro-patterns.** (*a*) Neurons imaged at $t$ = 16 hrs after plating. (*b*) Neurons imaged at $t$ = 32 hrs after plating. (*c*) Neurons imaged at $t$ = 64 hrs after plating. (*d*) Neurons imaged at $t$ = 80 hrs after plating. The scale bar is 15 μm in (*a*) and 20 μm in (*b-d*).

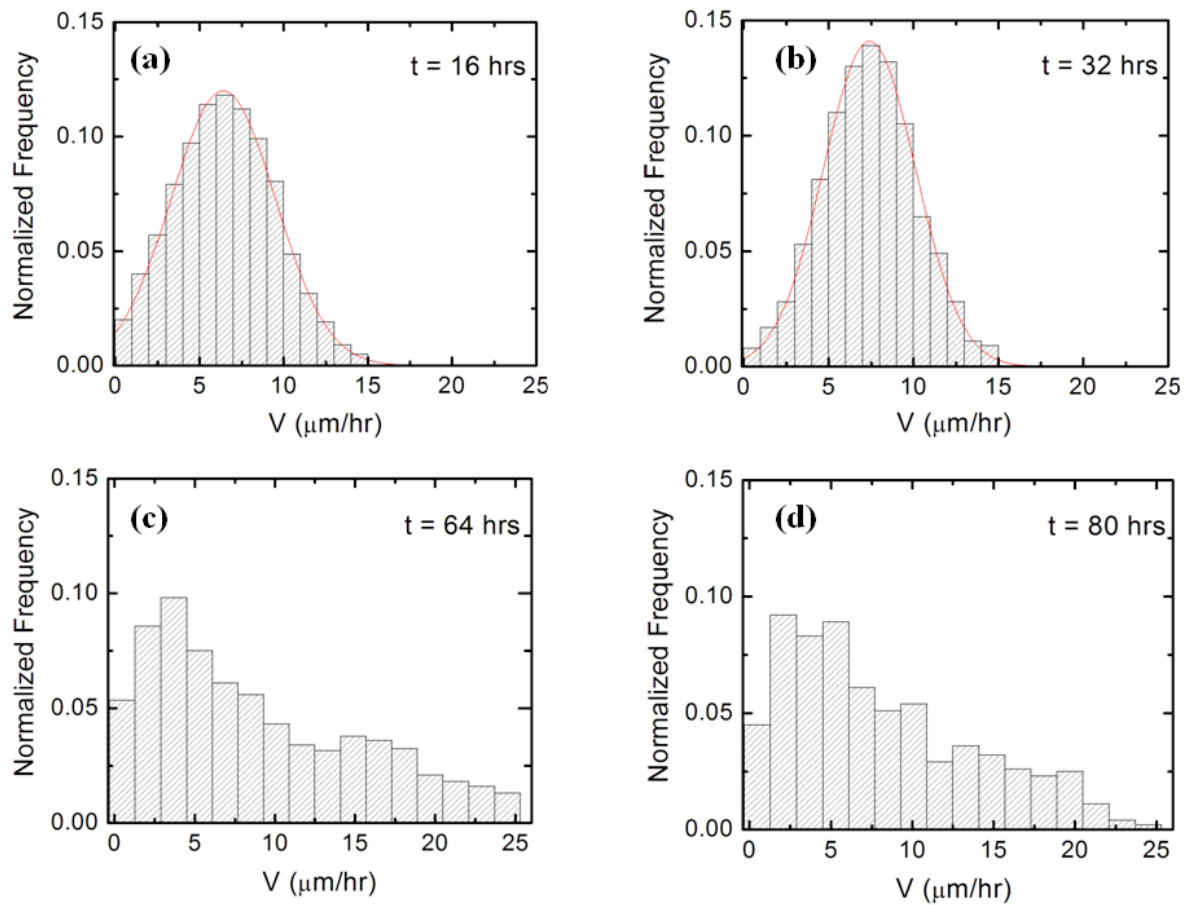

**FIGURE S2. Examples of normalized speed distributions for growth cones measured on PDMS substrates.** (*a*) Speed distribution for N = 182 different growth cones, measured at *t* = 16 hrs after plating. The continuous red curve represents fit with the Gaussian distribution given by Eq. 6. (*b*) Speed distribution for N = 195 different growth cones measured at *t* = 32 hrs after plating. The continuous red curve represents fit with the Gaussian distribution given by Eq. 6. (*c*) Speed distribution for N = 179 different growth cones measured at *t* = 64 hrs after plating. (*d*) Speed distribution for N = 168 different growth cones measured at *t* = 80 hrs after plating.

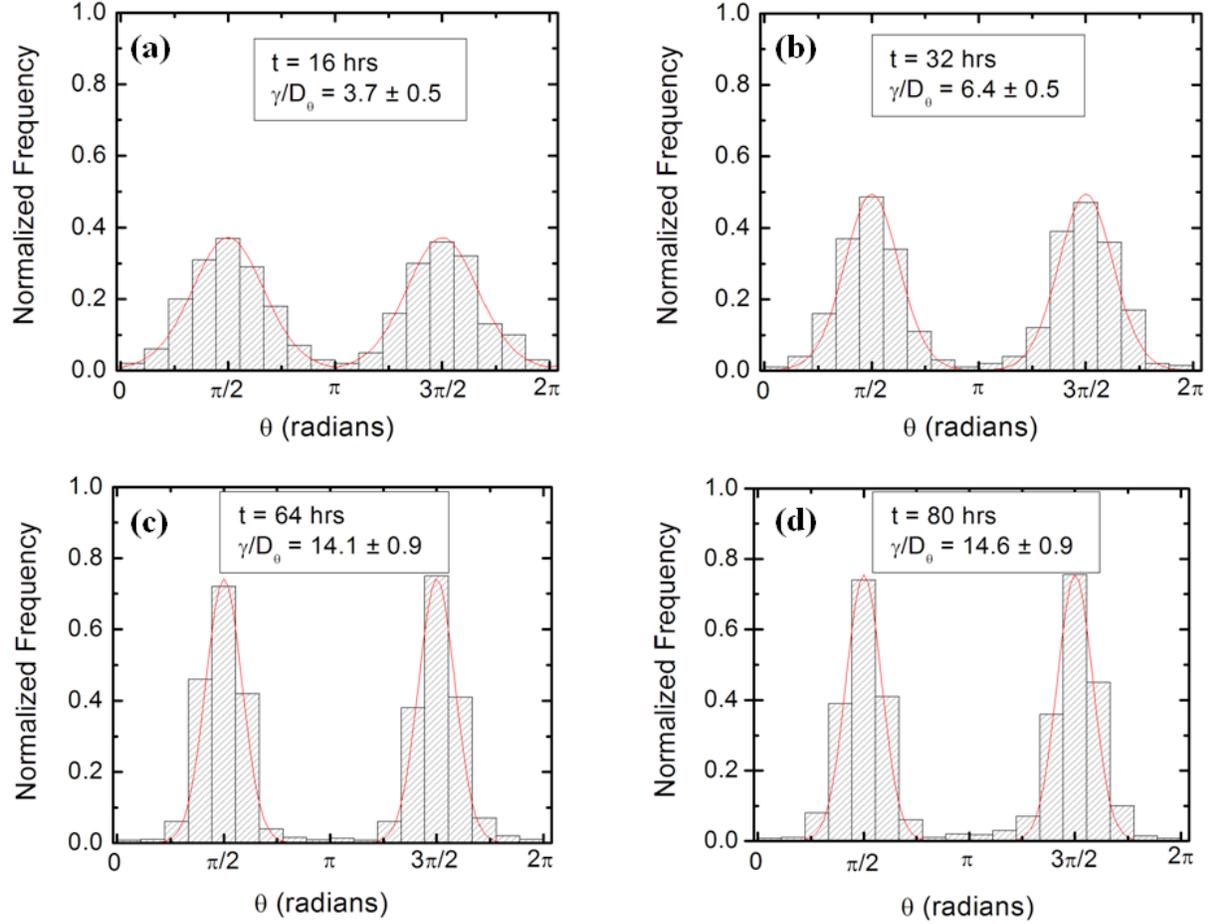

**FIGURE S3. Examples of normalized experimental angular distributions for axonal growth on patterned PDMS substrates.** The vertical axis (labeled Normalized Frequency) represents the ratio between the number of axonal segments growing in a given direction and the total number N of axon segments measured at a given time point $t$. Each axonal segment is of 20 μm in length (see Data Analysis section). (*a*) Data for N = 1877 different axon segments obtained at $t$ = 16 hrs after plating. (*b*) Data for N = 2383 different axon segments obtained at $t$ = 32 hrs after plating. (*c*) Data for N = 2537 different axon segments obtained at $t$ = 64 hrs after plating. (*d*) Data for N = 2903 different axon segments obtained at $t$ = 80 hrs after plating. The axons display strong directional alignment along the surface patterns (peaks at $\theta = \pi/2$ and $\theta = 3\pi/2$), with the degree of alignment (sharpness of the distribution) increasing with time. The continuous red curves in each figure represents fit to the data points using Eq. 14. The data fit gives the ratio $\gamma_\theta / D_\theta$ between the deterministic torque and the diffusion coefficient for the angular motion, at each time point (see text).

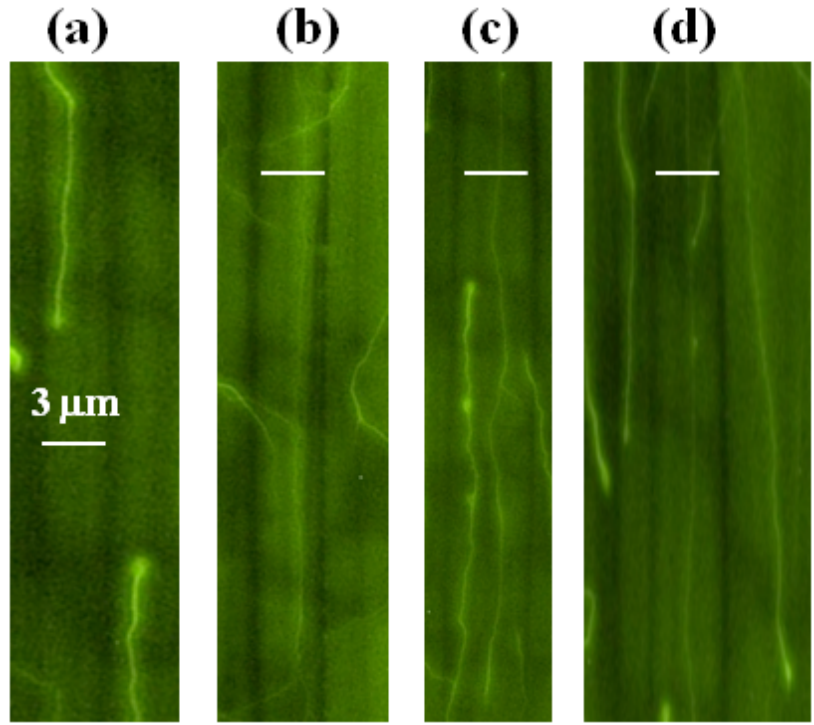

**FIGURE S4. Examples of fluorescence images showing the position of axons with respect to the patterns.** The images have been taken using the high magnification objective (60x) of the Nikon Eclipse Ti microscope, at different locations on 2 different substrates. The images show the fluorescently labeled microtubules (green), i.e. the C domain (see ref. [1]) inside the axons. The microtubules are labeled using Tubulin Tracker Green (see main text). The position of the micro-patterned troughs is shown by the vertical black lines. The 3μm white scale bar shows the distance between two adjacent troughs, and it has the same size for all images. The images show that the axons are located on the ridges of the patterns. The position of the ridges and troughs has been verified using AFM (images similar to the one shown in Fig. 2).

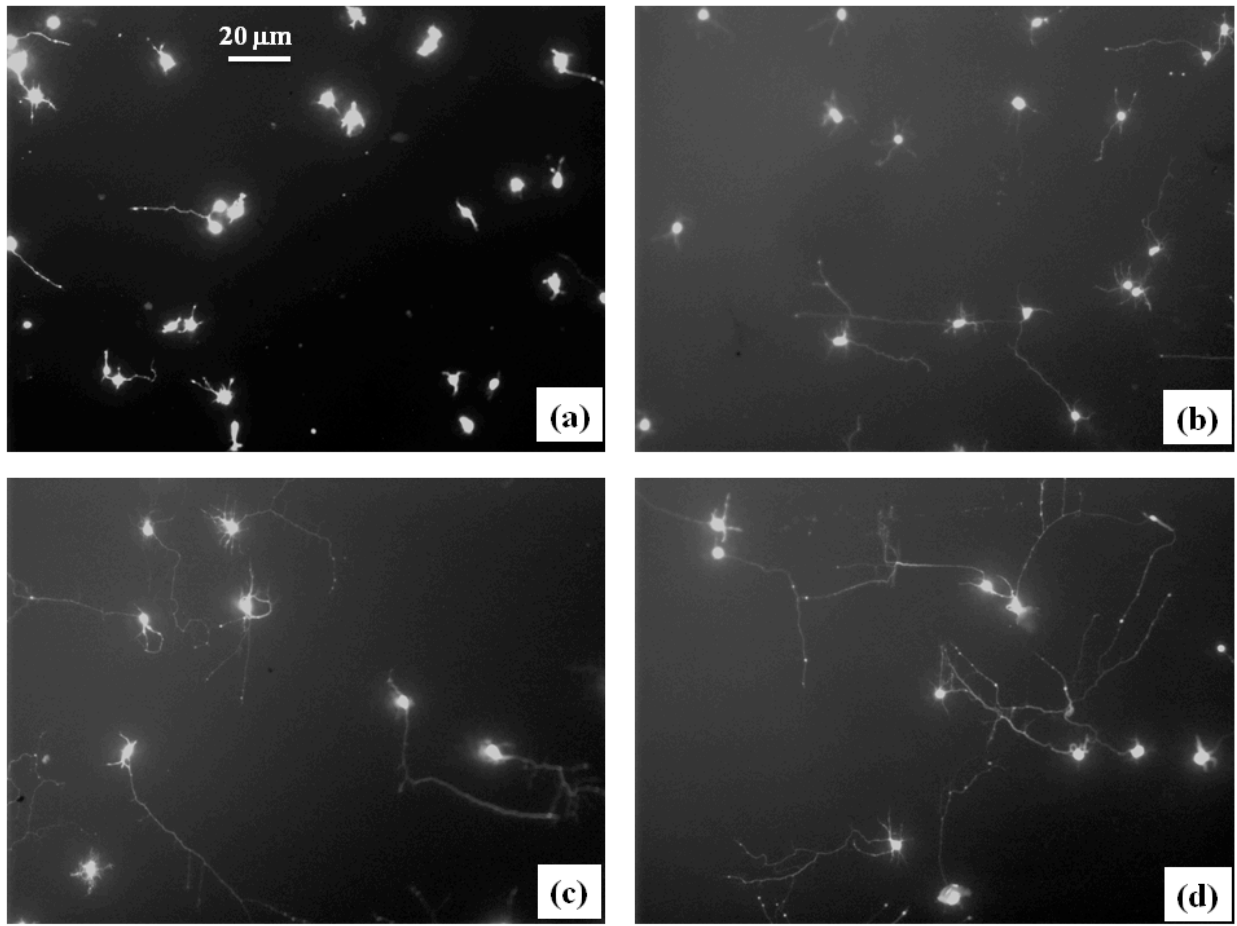

**FIGURE S5. Examples of cultured cortical neurons on flat (un-patterned) PDMS surfaces coated with PDL.** (*a*) Neurons imaged at $t = 6$ hrs after plating. (*b*) Neurons imaged at $t = 24$ hrs after plating. (*c*) Neurons imaged at $t = 48$ hrs after plating. (*d*) Neurons imaged at $t = 72$ hrs after plating. The scale bar is 20 μm in all images.

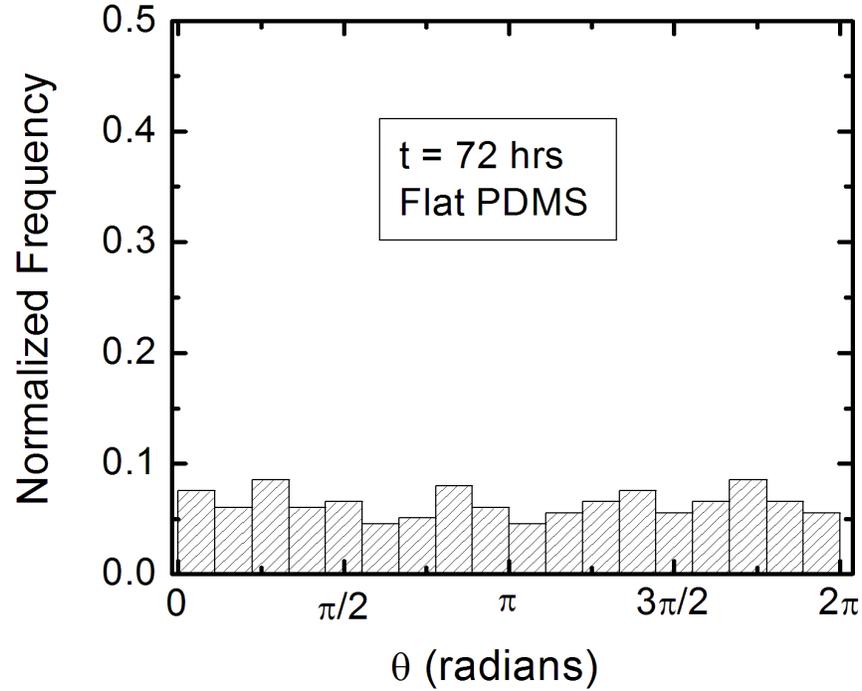

**FIGURE S6. Example of normalized experimental angular distributions for axonal growth on un-patterned PDMS substrates measured at *t = 72 hrs* after plating.** The vertical axis (labeled Normalized Frequency) represents the ratio between the number of axonal segments growing in a given direction and the total number N of axon segments measured. Each axonal segment is of 20 μm in length (see Data Analysis section). The data was taken for N = 1020 different axon segments measured at $t$ = 72 hrs after plating. The angular distribution demonstrates that there is no axonal alignment, in contrast to the case of neuronal growth on patterned PDMS (Fig. 5 and Fig. S3).

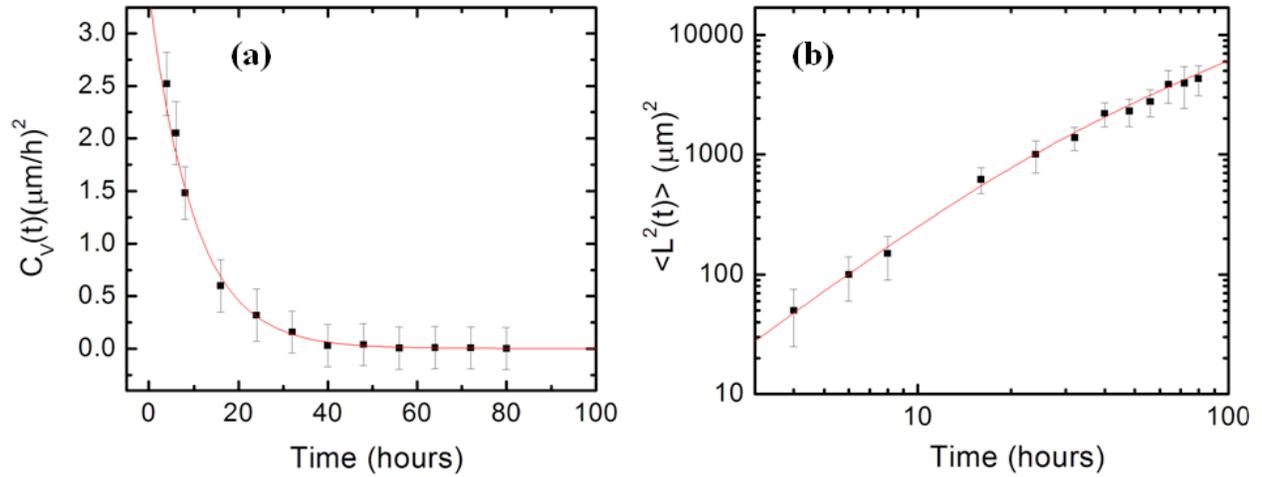

**FIGURE S7. Variation of the velocity autocorrelation function and axonal mean square length with time for neurons cultured on un-patterned PDMS surfaces.** (*a*) Data points: experimentally measured velocity autocorrelation function vs. time. The continuous red curve represents the fit of the data points with the prediction of the theoretical model based on the Ornstein-Uhlenbeck process (Eq. 7). *(b)* log-log plot of axonal mean square length vs. time. The continuous red curve represents the fit to the data with Eq. 8 (prediction of the theoretical model based on the Ornstein-Uhlenbeck process). Each data point in (*a*) and (*b*) was obtained by measuring between N = 70 and N = 137 different axons (corresponding to 3-6 different fluorescent images per time data point). Error bars in both figures indicate the standard error of the mean. The fit of the data in with Eq. 7 for (*a*), and Eq. 8 for (*b*) give the diffusion coefficient $D$ and the constant damping coefficient $\gamma$ of the Ornstein-Uhlenbeck process (see text).